\documentclass[12pt]{article}

\usepackage{graphicx}

\begin{document}

\title{Statistical Properties of E(5) and X(5) Symmetries}

\author{{ Jing Shu$^{1}$, Hong-bo Jia$^{1}$, Yu-xin Liu$^{1,2,3,
4,}$\thanks{Corresponding author} }\\[3mm]
\normalsize{$^1$ Department of Physics, Peking University, Beijing 100871, China}\\
\normalsize{$^2$ The Key Laboratory of Heavy Ion Physics of the
Chinese Ministry of Education, }\\
\normalsize{Peking University, Beijing 100871, China}\\
\normalsize{$^3$ Institute of Theoretical Physics, Academia
Sinica,
Beijing 100080, China} \\
\normalsize{$^4$ Center of Theoretical Nuclear Physics, National
Laboratory of}\\ \normalsize{ Heavy Ion Accelerator, Lanzhou
730000, China} }

\maketitle

\begin{abstract}
We study the energy level statistics of the states in E(5) and
X(5) dynamical symmetries. The calculated results indicate that
the statistics of E(5) symmetry is regular and follows Poisson
statistics, while that of X(5) symmetry involves two maxima in the
nearest neighbor level spacing distribution $P(s)$ and the
$\Delta_{3}$ statistics follows the GOE statistics. It provides an
evidence that the X(5) symmetry is at the critical point
exhibiting competing degrees of freedom.

\end{abstract}

\bigskip


\newpage

\parindent=20pt

\baselineskip=24pt

The nature of ``shape" phase transitions in finite quantal system
is a fundamental issue and has been the subject of many
investigations. In the last years, there have been many preeminent
works concerning this subject in nucleus, including both
theoretical calculations\cite{IZC98, CKZ99, JCD99, ZCZC99} and
experimental results\cite{CWRZB97, Zetal99}. Particularly, two
symmetries at the critical points of shape phase transition,
called E(5) and X(5), have recently been proposed\cite{Iac00,
Iac01}. It is a major breakthrough in the study of critical point
behavior of nucleus undergoing a shape phase transition because it
gives us simple analytic solutions while historically we have to
resort to numerical calculations to describe such nucleus.

Empirical examples of such kind of symmetries have been found soon
after they were put forward. For instances, the $^{134}$Ba,
$^{102}$Pd, $^{108}$Pd, Ru isotopes are proposed to be the ones in
E(5) symmetry\cite{CZ00, ZL02, FAA02, Zam02} and the $^{152}$Sm,
$^{150}$Nd, $^{104}$Mo in X(5) symmetry\cite{CZ01, KAC02, BB02}.
These examples show that the obtained new symmetries represent a
helpful theoretical tool which could well describe the structure
of the realistic nucleus at the critical point of a shape
transition, in a sense complementing the role played by the
dynamical symmetries of the Interacting Boson
Model(IBM)\cite{AI7389, IA87}. However, there remain some
interesting aspects on the statistical properties of the E(5) and
X(5) symmetries. (1) Whether the nuclear system displays ordered
or chaotic spectrum in the new symmetries? (2) Does the statistics
give a strong evidence to support that the obtained new symmetries
can well describe the nucleus at the critical point in a shape
transition? (3) The new symmetries are based on the particular
potential amenable to analytic descriptions which well approximate
to the ``true" potential. Does the simplification of the potential
eliminates part of the statistical properties of the realistic
nucleus at the critical point? Or more concretely, are the results
consistent with the statistical properties obtained by Alhassid
and collaborates\cite{AN92,AW91,ANW90} in the framework of IBM?
Therefore it is crucially important to obtain the statistics of
the new symmetries and answer the above opening questions. In this
sense, we explore the statistical properties of E(5) and X(5)
symmetries in the paper.

We start from the Bohr Hamiltonian\cite{B52}
\begin{equation}
H\! = \! -\! \frac{\hbar ^2}{ 2 B} \Big[
\frac{1}{\beta^{4}}\frac{\partial} {\partial \beta} \beta ^4
\frac{\partial}{\partial \beta} + \frac{1}{\beta ^2 \sin{ 3
\gamma}} \frac{\partial}{\partial \gamma} \sin { 3 \gamma}
\frac{\partial}{\partial \gamma} - \frac{1}{4 \beta ^2}
\sum_{\kappa}{\frac{Q_{\kappa}^2}{\sin ^{2} (\gamma - \frac{2}{3}
\pi \kappa) }} \Big] + V(\beta , \gamma ) .
\end{equation}
In the E(5) symmetry, the potential $V(\beta,\gamma)$ depends only
on the deformation parameter $\beta$ and it can be written as
$V(\beta,\gamma) = U(\beta)$. Considering the case, in which the
potential is simplified as an infinite well, one has the
eigenvalues\cite{Iac00}
\begin{equation}
E_{\xi, \tau} = \frac{ \hbar ^2}{2 B} \Big(\frac{\chi_{\xi ,
\tau}}{\beta _{\mbox{w}} }\Big)^{2} \, ,
\end{equation}
where $B$ is a parameter, $\chi_{\xi , \tau}$ is the $\xi$th zero
of the Bessel function $J_{\nu}(z)$, and the order of the Bessel
function is $\nu = \tau + \frac{3}{2}$ with $\tau$ being the
quantum number similar to the irreducible representation of O(5)
group in the IBM, $\beta_{\mbox{w}}$ is the width of the well.

In the X(5) symmetry, the potential was supposed to be a square
well in the variable $\beta$ and a harmonic oscillator in $\gamma$
with no $\beta-\gamma$ couplings, which is a good approximation in
the U(5)-SU(3) shape transition. In this case, the eigenvalues are
given as\cite{Iac01}
\begin{equation}
E(s, L, n_\gamma, K, M) =  E_{\beta} + E_{\gamma} \, ,
\end{equation}
where
\begin{equation}
E_{\beta} =  {\Big(\frac{\chi _{s, L}}{\beta_{\mbox{w}}}\Big)}^{2}
\,
\end{equation}
and
\begin{equation}
E_{\gamma} =
\frac{3a}{\sqrt{\langle\beta^{2}\rangle}}(n_{\gamma}+1) -
\frac{4}{3} \frac{(K/2)^{2}}{\langle\beta^{2}\rangle} \, .
\end{equation}
The $\chi _{s, L}$ is the $s$th zero of Bessel function
$J_{\nu}(z)$ of (irrational) order $ \nu = [J(J+1)/3+9/4]^{1/2}$,
$L$ is the total angular momentum (with projections $K$ on the
symmetry axis and $M$ on the quantization axis in the laboratory
frame). $a$ is a parameter related to the interaction strength of
the potential $U_{\gamma} = (3a)^{2}\gamma^{2}/2$ and $n_\gamma$
is the number of $\gamma$-vibration quanta, while
$\beta_{\mbox{w}}$ is the same as that in E(5) symmetry and the
$\langle\beta^{2}\rangle$ is the average of $\beta^{2}$ over the
wave functions. Combining all variables, one obtains the most
general expression\cite{Iac01}
\begin{equation}
E(s, L, n_\gamma, K, M) =  E_{0} + B(\chi _{s, L})^{2} +A
n_{\gamma} + CK^{2} \, ,
\end{equation}
where
\begin{equation}
A = \frac{3a}{\sqrt{\langle\beta^{2}\rangle}} \,
\end{equation}
and
\begin{equation}
B = \Big(\frac{1}{\beta_{\mbox{w}}}\Big)^{2} \, .
\end{equation}
Before analyzing the energy level statistics, we estimate the
relative value of the parameters $A/B$ in the light of the ``true"
potential given in the Fig.1. of Ref.\cite{Iac01}. We can see from
the figure that the width of the infinite well to approximate the
``true" potential is $\beta_{\mbox{w}} \approx 0.8$, and most of
the quantum states can thus be restricted in $U_{\beta}<0.5$
(arbitrary unit). If we assume that the total energy of the
quantum states are conserved in $E=0.5$, the quantum states of the
$\gamma$ degree are then mainly restricted in $U_{\gamma}<0.5$, as
the potential is completely decoupled into two separate parts in
$\beta$ and $\gamma$ respectively. Since the X(5) symmetry is
obtained around $\gamma=0$\cite{Iac01}, the corresponding
restriction in $\gamma$ by the total energy must be very small.
Taking this small value of $\gamma$ as 0.1, we then work out that
the strength of the two dimensional oscillator $(3a)$ must be more
than 10 according to the formulation $U_{\gamma} =
(3a)^{2}\gamma^{2}/2$. With Eqs.(7) and (8), we obtain that the
parameter $A$ must be more than 25 if we take
$\sqrt{\langle\beta^{2}\rangle}$ as 0.4 and the parameter
$B=1.56$. Consequently we get the relative value $A/B \geq 16$.

Our estimate is consistent with the recent empirical example
$^{104}$Mo\cite{BB02} of X(5) symmetry, where the nucleus presents
the X(5) pattern not only in its ground state band, but also in
its low-lying $n_{\gamma} = 1,2; K=2n_{\gamma}$ bands. From the
level scheme given in the Fig.2. of Ref.\cite{BB02}, we can obtain
the empirical values of the parameters, $A\approx800$ and
$B\approx30$ (in keV). The relative value is clearly in the region
of $A/B \geq 16$ as we have proposed. Such a consistence supports
that the assumption to obtain X(5) symmetry around $\gamma=0$ in
the Ref.\cite{Iac01} is quite reasonable since in fact, most of
the quantum states in the realistic nucleus at the critical point
of the transition from spherical to axially-deformed shape are
restricted in $\gamma<0.1$.

To analyze the energy level statistics in the dynamical
symmetries, we take the following process. At first with Eqs.(2)
and (6), we calculate the energy levels of the nucleus in E(5) and
X(5) symmetries, respectively. The parameter $\xi$ and $\tau$ in
E(5) symmetry are truncated manually at a maximal number
$\xi^{(m)}$ and $\tau^{(m)}$. We also truncated $s$ and $n_\gamma$
at $s^{(m)}$ and $n_{\gamma}^{(m)}$ respectively in X(5) symmetry.
After the spectrum $\{E_i\}$ has been determined, it is necessary
to separate the smoothed average part whose behavior is
nonuniversal and can not be described by random-matrix
theory(RMT)\cite{meta67}. To do so we count the number of the
levels below $E$ so that one can define a staircase function
$N(E)$ of the spectrum (see for example Ref.\cite{AN92})
$$\displaylines{\hspace*{1cm}
N(E)=N_{av}(E)+N_{fluct}(E) \, . \hfill{(10)} \cr }
$$
Then we fix the $N_{av}(E_i)$ semiclassically by taking a smooth
polynomial function to fit the staircase function $N(E)$. We
obtain finally the unfolded spectrum with the mapping
$$\displaylines{\hspace*{1cm}
\{\widetilde{E}_i \}=N(E_i) \, . \hfill{(11)} \cr }
$$
This unfolded level sequence $\{\widetilde{E}_i\}$ is obviously
dimensionless and has a constant average spacing of 1, but the
actual spacings exhibit frequently strong fluctuation.

We have used two statistical measures to determine the fluctuation
properties of the unfolded levels: the nearest neighbor level
spacings distribution(NSD) $P(S)$ and the spectral rigidity
$\Delta_3(L)$. The nearest neighbor level spacing is defined as
$S_i=(\widetilde{E}_{i+1})-(\widetilde{E}_i)$. The distribution
$P(S)$ is defined as that $P(S)dS$ is the probability for the
$S_i$ to lie within the infinitesimal interval $[S, S+dS]$. For a
regular system, it is expected to behave like the Poisson
statistics
$$\displaylines{\hspace*{1cm}
P(S)=e^{-S} \, , \hfill{(12)} \cr }
$$
whereas if the system is chaotic, one expects to obtain the Wigner
distribution
$$\displaylines{\hspace*{1cm}
P(S)=(\pi/2)S \ \textrm{exp}(-{\pi}S^{2}/4) \, , \hfill{(13)} \cr
}
$$
which is consistent with the GOE
statistics\cite{meta67,Brod81,port65}. With the Brody parameter
$\omega$ in the Brody distribution
$$\displaylines{\hspace*{1cm}
P_{\omega}(S)=\alpha(1+\omega)S^{\omega}
\textrm{exp}({-\alpha}S^{(1+\omega)}) \, , \hfill{(14)} \cr }
$$
where
$$\displaylines{\hspace*{1cm}
\alpha=\Gamma[(2+\omega)/(1+\omega)]^{1/2} \, \hfill{(15)} \cr }
$$
and $\Gamma[x]$ is the $\Gamma$ function, the transition from
regularity to chaos can be measured with the Brody parameter
$\omega$ quantitatively. It is evident that $\omega=1$ corresponds
to the GOE distribution, while $\omega=0$ to the Poisson
distribution. A value $0<\omega<1$ means an interplay between the
regular and the chaotic.

As to the spectral rigidity $\Delta_{3}(L)$, it is defined as
$$\displaylines{\hspace*{1cm}
\Delta_{3}(L)={\Bigg\langle}min_{A,B}\frac{1}{L}\int_{-L/2}^{L/2}
[N(x)-Ax-B]^{2}dx{\Bigg\rangle} \, , \hfill{(16)} \cr }
$$
where $N(x)$ is the staircase function of a unfolded spectrum in
the interval $[-L/2,x]$. The minimum is taken with respect to the
parameters $A$ and $B$. The average denoted by
$\langle\cdots\rangle$ is taken over a suitable energy interval
over $x$. Thus from this definition $\Delta_{3}(L)$ is the local
average least square deviation of the staircase function $N(x)$
from the best fitting straight line. For the GOE the expected
value of $\Delta_{3}(L)$ can only be evaluated numerically, but it
approaches the value
$$\displaylines{\hspace*{1cm}
\Delta_{3}(L)\cong(lnL-0.0687)/\pi^2 \, \hfill{(17)} \cr }
$$
for large $L$. And for Poisson statistics
$$\displaylines{\hspace*{1cm}
\Delta_{3}(L)=L/15 \, . \hfill{(18)} \cr }
$$

In our calculation of E(5) symmetry, all the degenerate states
caused by additional quantum number are considered just as one
single state\cite{SRJL02}. That is because we manually introduced
the additional quantum number to distinguish the degenerate states
with the same quantum number $\tau$. The degenerate states with
different additional quantum numbers are statistically
uncorrelated since no interaction is involved to link such states.
If we do not remove the degeneracy, we would take the mixed
ensembles (with different additional quantum numbers which are
good quantum numbers in the dynamical symmetries) into
consideration and obtain the over-Poisson distribution in
practical calculations since nearly 1/4 of the spacings of all are
zero. In order to evaluate the fluctuations of the energy levels
in one pure ensemble, these large amount of the statistically
uncorrelated states should be removed.

At first, we analyze the energy levels given in Eqs.(2) for the
E(5) dynamical symmetry. The numerical results of the NSD and
$\Delta_3$ statistics show that the statistical feature does not
depend on the angular momentum obviously. We then illustrate only
those of the states with low spin-parity $J^\pi=0^+$ but different
sets of truncated numbers $\xi^{(m)}$ and $\tau^{(m)}$ of E(5)
symmetry in Fig.1 (a), (b) and (c). Meanwhile the Brody parameter
$\omega$ of the NSD\cite{Brod81} is also evaluated and listed in
the figures. The numerical results show that the statistics of the
E(5) symmetry is quite close to Poisson statistics. When the
number of the energy levels involved in the statistics (denoted by
$\xi^{(m)}$ and $\tau^{(m)}$) increases, the statistics becomes
almost exactly the Poisson-type. It indicates that the E(5)
symmetry is a dynamical symmetry which is completely integrable in
classical limit.

We then analyze the statistics in the X(5) symmetry. We can easily
find that the energy eigenvalues in the X(5) symmetry contain two
terms, one is the square of the zeros of the Bessel function in
$\beta$, another is the solution of two dimensional oscillator
function in $\gamma$. In order to display how the relative
strength of the two terms affects the energy level statistics, we
calculate the statistics in different values of the relative
strength $A/B$, the results are displayed in Fig.2. All of the
relative strength $A/B$ satisfy the condition $A/B\geq16$ as we
estimated above. It is apparent that when the parameter $A/B$ is
not very large, two maxima\cite{SRJL02} appear in the NSD
statistics (see the left part of Fig.2. (a), (b)), which is rather
different from the fluctuation properties in other cases except
for those in the U(5) dynamical symmetry with collective
backbending\cite{SRJL02, Long97}. In theoretical point of view,
the X(5) symmetry describes the critical point nuclei in a
spherical to axially deformed shape phase transition analytically.
The structure of its energy levels must be not the same
everywhere. In fact, the term $B(\chi_{s,L})^{2}$ is approximately
a rotational term of $s$, whereas the term $An_{\gamma}$ is a
vibrational term of $n_{\gamma}$. The simultaneous appearance of
these two terms induces a competition and makes the ensemble with
the same angular momentum $L$ and its projection on the symmetry
axis $K$ involves two sequences, one of which is in vibration,
another is in rotation. When we unfold the above spectrum
$\{E_{i}(L)\}$, the two sequences are normalized with an unique
total average spacing, and their maxima in the spacing
distributions are no longer the same, consequently two maxima
emerge because of the uniform unfolding procedure. The same
mechanism and results can be found in U(5) dynamical symmetry with
collective backbending, where the the structure of the yrast band
changes from the U(5) vibrational states to the rotational states
with full d-boson number $n_{d}=N$, which implies that a
transition of collective motion mode may happen as the angular
momentum increases. In this statistical point of view, the X(5)
symmetry describes well the critical point behavior of nucleus
undergoing a shape transition from U(5) dynamical symmetry to
SU(3) dynamical symmetry which involves competing degrees of
freedom.

In the $\Delta_{3}$ statistics when $A$ is not very large (see the
right part of Fig.2. (a), (b)), it is clearly that the statistics
follows the GOE statistics, which suggests that the competing
strengthes in the $\beta$ and $\gamma$ degrees can cause the onset
of chaos in X(5) symmetry. The case that the degree of chaoticity
reaches its maximal when different competing strengthes coexist
and are comparable can be also found in other works (see for
example Ref.\cite{SRJL02, AV92}).

Looking over Fig.2.(c) and (d), one can recognize that when we
increase the strength of the oscillator in $\gamma$ degree of
freedom, which corresponds to the parameter $A$, the two maxima in
the NSD statistics soon vanishes and both the NSD and $\Delta_{3}$
statistics become close to the Poisson statistics. It indicates
that only the vibrational mode plays a dominant role in the system
and the competition mentioned above does not exist. If we go to
extremes to the case when the vibrational strength $3a
\rightarrow\infty$, the $\gamma$ degree of freedom no longer
exists. The X(5) symmetry becomes the E(5) symmetry in a
four-dimensional space, and its statistics is certainly regular.

The results obtained above are consistent with Alhassid and
collaborates' work\cite{AW91}. The E(5) symmetry lies in the
transitional region between a vibration to a $\gamma$-unstable
rotation, which is regular both classically and quantum
mechanically in the work. It has been known that the Hamiltonian
in the consistent $Q$ formalism\cite{WC82} of the IBM can be given
as
\begin{equation}
H = \epsilon \hat{n_{d}} - \kappa \hat{Q}^{2}(\chi)\cdot
\hat{Q}^{2}(\chi) \, ,
\end{equation}
where
$$\displaylines{\hspace*{5cm}
\hat{Q}^{2}_{\mu}(\chi) = (s^{\dag} \tilde{d} + d^{\dag}
s)^{2}_{\mu} + \chi (d^{\dag} \tilde{d} ) ^{2}_{\mu} \, . \hfill{}
\cr }$$
 By changing $\epsilon/\kappa$ under restriction
$\chi = -\sqrt{7}/2$, one can describe the U(5)-SU(3) transition
with this Hamiltonian. To discuss the shape phase transition and
relate it with X(5) symmetry, we reparametrize the ration
$\epsilon/\kappa$ as $(1-\zeta)/\zeta$. From the Ref.\cite{IZC98}
we know that the flat potential near the critical point could be
obtained when the parameter $\zeta\in$(0.025, 0.029)\cite{IZC98}
in case that the boson number $N = 10$. We then transfer the
parameter $\zeta$ into parameter $\eta$ used in Alhassid and
collaborates' work\cite{AW91}. The relation is
$$\displaylines{\hspace*{5cm}
\eta = \frac{1-\zeta}{1-\zeta+N\zeta} \, , \hfill{} \cr }$$ and
the value of $\eta$ near the critical point is $0.770 \sim 0.796$.
It is clearly that the critical point which corresponds to the
X(5) symmetry lies in the most chaotic region(see the Fig.1. in
Ref\cite{AW91}).

Finally, we turn to the question we have raised in the beginning.
In X(5) symmetry the nucleus exhibits chaotic spectrum, which is
probably caused by the comparable competing strengthes in the
$\beta$ and $\gamma$ degrees. The two maxima appearing in the NSD
statistics suggest that the spectrum splits into two different
sequences, one is rotational and the other is vibrational. Such a
behavior illustrates that the nucleus in X(5) symmetry is at the
critical point of the transition from spherical to
axially-deformed shape with competing degrees of freedom. While
the statistics in E(5) symmetry provides no evidence for such a
drastic competition in the critical shape. That might be because
the E(5) symmetry is a critical symmetry corresponding to an
isolated point of second order phase transition. Very recently, it
has been pointed out that such a second-order phase transition
located at the triple-point of the three phases\cite{JCCHLW82}.
The shape coexistence region shrinks\cite{IZC98} as the parameter
$\chi$ in the Casten's triangle changes from $-\sqrt{7}/2$ to 0.
In deed, as a general rule, phase coexistence can be found only in
first-order phase transitions\cite{LL01}. There is no shape
coexistence in E(5) symmetry and the spectrum does not split into
two sequences, then no competition exists and its statistics is
uniform(no two maxima in the NSD statistics). The common
subalgebra O(5) in the U(5)$-$O(6) transition makes the statistics
regular in this region and the ordered spectrum of nucleus in E(5)
symmetry confirms the past work\cite{AW91}. It is worth to mention
that the bifurcation effect(number of maxima more than one) in the
NSD statistics might be a useful tool to determine whether the
system is at the critical point of a first order shape transition
or shape coexistence which involves competing degrees of freedom.

In summary, we have calculated the statistics of E(5) and X(5)
symmetries. The statistics of X(5) symmetry involves two maxima in
the NSD statistics and the $\Delta_3$ statistics follows the GOE
statistics, which provides an evidence that the X(5) symmetry is
at the critical point exhibits competing degrees of freedom. The
statistics of E(5) symmetry is regular. That is probably caused by
the common subgroup O(5) in the U(5)$-$O(6) transition. Since the
E(5) symmetry corresponds to the second order shape phase
transition with no shape coexistence, then no such competition
exists as in the X(5) symmetry.

\vspace*{8mm}

This work is supported by the National Natural Science Foundation
of China under the contract No. 19875001, 10075002, and 10135030.
The authors (J. S. and H. J.) thank the support from the Taozhao
and Tsung-zheng Foundation at Peking University, respectively. The
other author (Y. L.) acknowledges the support by the Foundation
for University Key Teacher by the Ministry of Education, China and
the Founds of the Key Laboratory of Heavy Ion Physics at Peking
University, Ministry of Education, China, too.

\newpage

\begin{figure}
\begin{center}
\includegraphics[width=0.32\textwidth,angle=0]{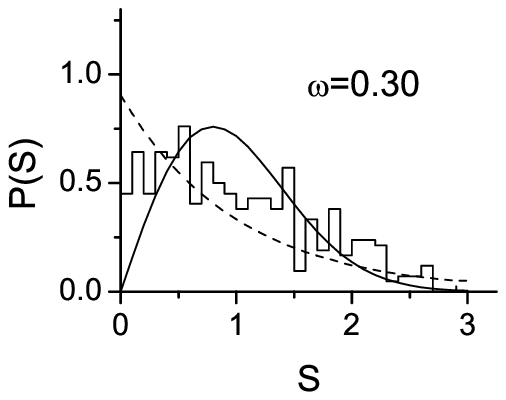}
\includegraphics[width=0.32\textwidth,angle=0]{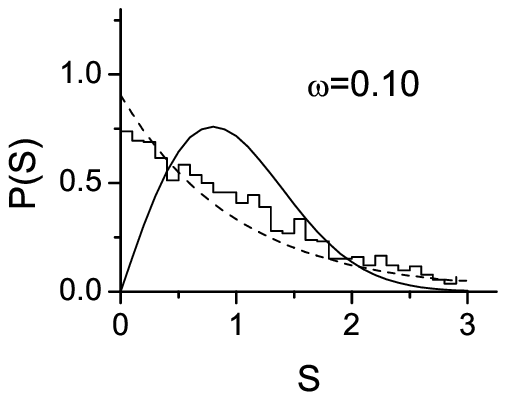}
\includegraphics[width=0.32\textwidth,angle=0]{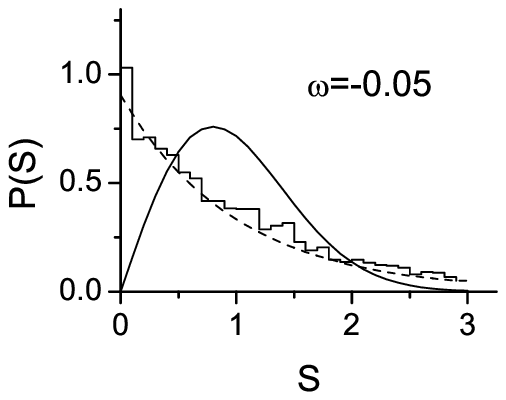}
\includegraphics[width=0.32\textwidth,angle=0]{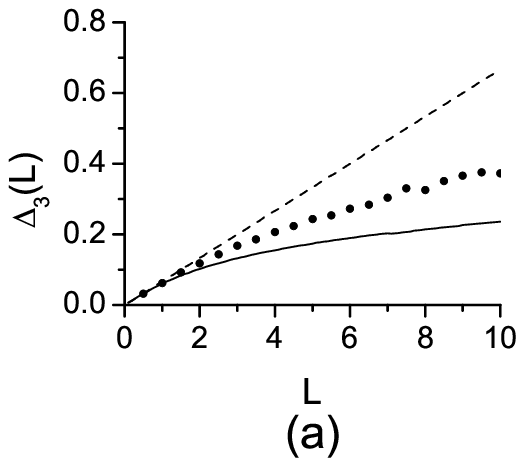}
\includegraphics[width=0.32\textwidth,angle=0]{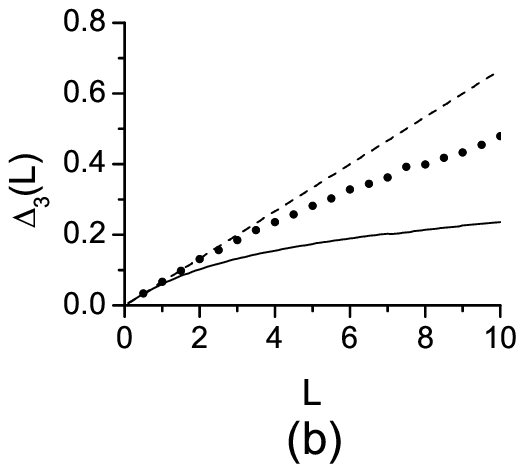}
\includegraphics[width=0.32\textwidth,angle=0]{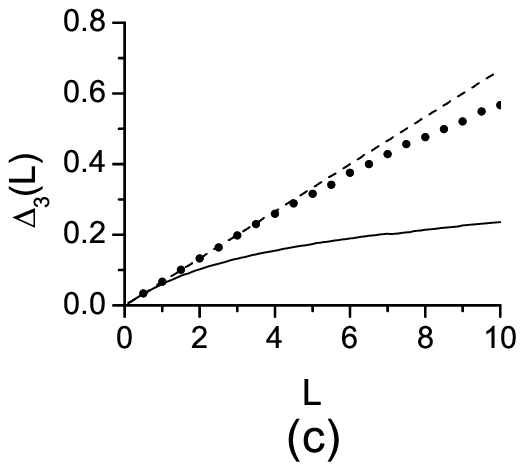}
\bigskip
\caption{Comparison of energy level statistics of the states
$J^\pi=0^+$ in E(5) symmetry with different numbers of energy
levels: (a) for $\tau^{(m)}=20$, $\xi^{(m)}=20$, (b) for
$\tau^{(m)}=40$, $\xi^{(m)}=40$, and (c) for $\tau^{(m)}=80$,
$\xi^{(m)}=80$. In all figures, the solid lines and dashed lines
describe the GOE and Poisson statistics, respectively.}
\end{center}
\end{figure}

\newpage
\begin{figure}
\begin{center}
\includegraphics[width=0.34\textwidth,angle=0]{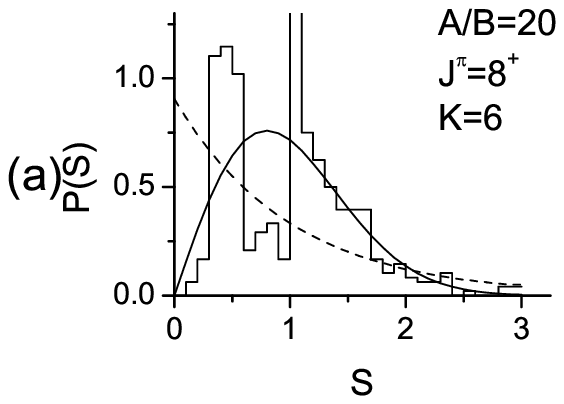}
\includegraphics[width=0.31\textwidth,angle=0]{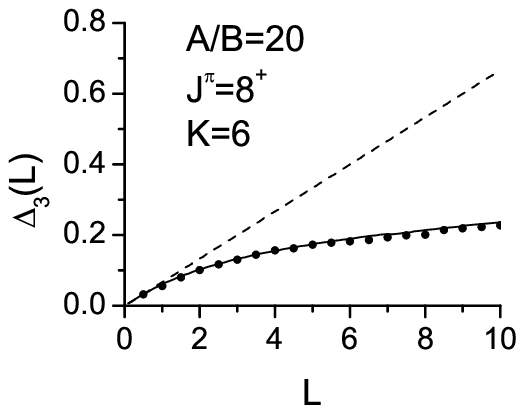}
\includegraphics[width=0.34\textwidth,angle=0]{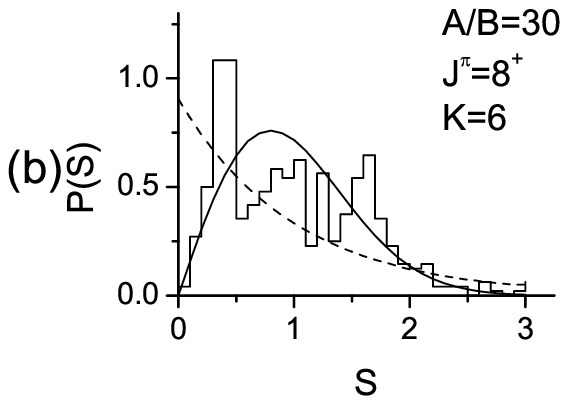}
\includegraphics[width=0.31\textwidth,angle=0]{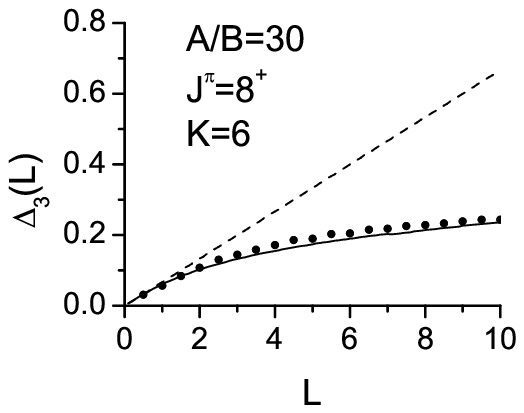}
\includegraphics[width=0.34\textwidth,angle=0]{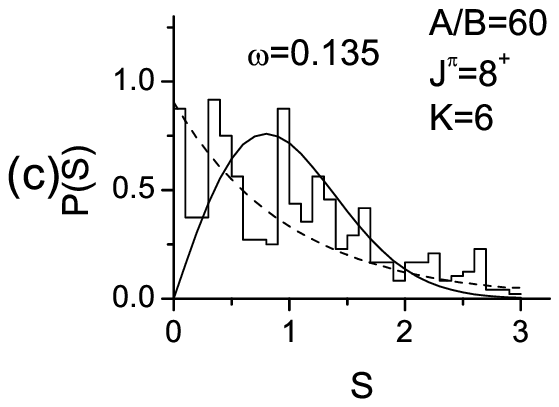}
\includegraphics[width=0.31\textwidth,angle=0]{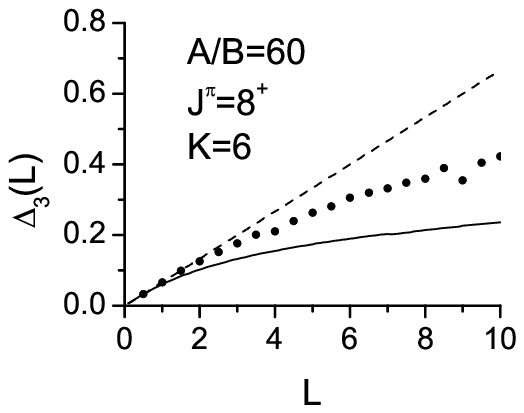}
\includegraphics[width=0.34\textwidth,angle=0]{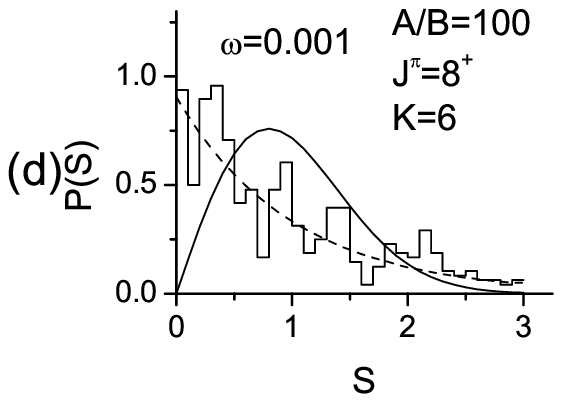}
\includegraphics[width=0.31\textwidth,angle=0]{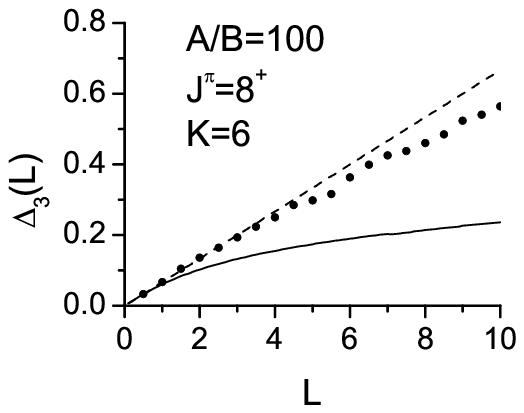}
\bigskip
\caption{Comparison of energy level statistics of the states
$J^\pi=8^+$ and $K=6$ in X(5) symmetry with the same manually
truncated quantum number $s^{(m)}=30$ and $n_{\gamma}^{(m)}=30$,
but different relative strength $A/B$: (a) for $A/B=20$, (b) for
$A/B=30$, (c) for $A/B=60$, and (d) for $A/B=100$.}
\end{center}
\end{figure}

\end{document}